\documentclass[%
reprint,
superscriptaddress,
nofootinbib,
amsmath,amssymb,
aps,
prb,
longbibliography
]{revtex4-2}

\usepackage{graphicx}

\usepackage{bm}
\usepackage{xcolor}

\usepackage[mathlines]{lineno}

\usepackage[colorlinks=true, linkcolor=blue,urlcolor=blue,citecolor=blue]{hyperref}
\usepackage{cleveref}

\usepackage[ignoreunlbld,norefs,nocites]{refcheck}

\begin{document}

\preprint{APS/123-QED}

\title{Effective and Floquet Hamiltonians for High Frequency Driving and Floquet-induced Heating in Quantum Spin Chains}

\author{Mayukh Bandyopadhyay}
\affiliation{%
Computational Quantum Many-Body Physics Lab, Department of Physics, Dr.\ B.\ R.\ Ambedkar National Institute of Technology, Jalandhar, Punjab - 144008, India}%

\author{Vinod Ashokan}
\email{ashokanv@nitj.ac.in}
\affiliation{Computational Quantum Many-Body Physics Lab, Department of Physics, Dr.\ B.\ R.\ Ambedkar National Institute of Technology, Jalandhar, Punjab - 144008, India}

\date{\today}
\begin{abstract}
We study the non-equilibrium dynamics of a disordered periodically driven quantum spin chain, with the competition between the interaction, disorder, and Floquet driving being of particular interest. We study dynamics of entanglement entropy, energy absorption to characterize dynamical regimes of the system whether it stays in the Floquet-MBL(many-body localization) region or thermalized region. Starting with a product state in the computational basis, followed by reduced density matrix which in turn gives rise to the entanglement entropy density. With the strength of the interaction, the transverse field, the parallel field, the disorder strength W and the driving frequency, we discover the distinct behaviors of fast delocalization and logarithmic entanglement growth and long-lasting memory of the initial state, indicative of localized or prethermal Floquet regimes. We observe that strong disorder arrests transport and enables slow entanglement dynamics, whereas strong driving frequency arrests energy absorption and creates a long-lived non-equilibrium state. Conversely, weak disorder or low driving frequency leads to delocalization. The outcomes show strong support for non-equilibrium phases in driven many-body systems.

\end{abstract}

 \maketitle


\section{Introduction}
Non-equilibrium physics of isolated quantum systems has received much interest in condensed matter physics. The recent development of ultracold atomic gases, trapped ions, and other highly controlled platforms for the study of quantum systems has led to the experimental realization of almost isolated many-body systems, thus inspiring a wide range of theoretical and experimental studies of their dynamics \cite{Greiner2002Collapse,MBPwithUltra,Blatt2012Quantum,Eisert2015Quantum}. Since these systems are isolated from external reservoirs, their quantum dynamics are governed by unitary time evolution according to the Schrödinger equation. A central issue is understanding the nature of rise or decay processes that emerge in such systems during unitary evolution \cite{ThermandItsmechgeneric,Phy,ChaosandQuantTherm,FromQuantChaosandEigenstateThermtoStat}. Thermalization in classical and quantum systems is an important phenomenon in statistical physics. Classically, the system, therefore, evolves to equilibrium configurations such that information about the initial state is lost from the macroscopic point of view \cite{nandkishore2015many,ThermandItsmechgeneric,PerDriErandMBL,ErgTheoStatMech}. Irreversibility of the reversible microscopic dynamics is discussed using probabilistic arguments, ergodicity, and the dominance of macrostates in phase space \cite{ErgTheoStatMech}. Thermalization in quantum systems, on the other hand, is more subtle, since the dynamics are controlled by unitary time evolution, leaving the quantum information of the initial state intact \cite{Phy,ThermandItsmechgeneric,ChaosandQuantTherm}. Thermalization is known to happen for generic interacting quantum systems, thanks to the Eigenstate Thermalization Hypothesis (ETH): local observables computed in individual many-body eigenstates are equal to those predicted by equilibrium statistical mechanics \cite{ChaosandQuantTherm,FromQuantChaosandEigenstateThermtoStat,Thermentg,Eisert2015Quantum}. In unitary evolution, two quantum systems evolve independently and, overall, they essentially behave like their own thermal baths, with the evolution of subsystems eventually bringing them to thermodynamic equilibrium while the whole system stays a pure quantum state \cite{ThermandItsmechgeneric,FromQuantChaosandEigenstateThermtoStat,nandkishore2015many}. Thus, at long times the local observables forget their initial state, and the behavior at long times can be well captured using the standard quantum statistical ensembles \cite{ThermandItsmechgeneric,FromQuantChaosandEigenstateThermtoStat,nandkishore2015many,ChaosandQuantTherm,Eisert2015Quantum}. It is through quantum entanglement and dephasing that the information is distributed nonlocally throughout the system as it maintains its global unitarity that the apparent irreversibility emerges \cite{Nonintg,Univslow,MBLandTherm,Colloquium}.

The paradigm of thermalization is nevertheless invalidated in the presence of strong disorder \cite{DisorderinQuantumManyBody}. In particular, at sub-cubic ratio of disorder to number of constituents, disordered interacting systems can fail to thermalize, leading to the phenomenon of many-body localization (MBL) \cite{nandkishore2015many}. This phase is a sign of the loss of ergodicity in isolated quantum systems; in these systems, the system is not an efficient thermal bath for its subsystems, and the ETH breaks down \cite{Locofinteractingfermionsathigh,MBLandTherm,nandkishore2015many,Univslow,Colloquium}. Unlike ergodic systems which will ultimately reach thermalization, many-body localized states retain information about their initial condition for long times and evade thermal equilibrium even in highly excited states \cite{Locofinteractingfermionsathigh,nandkishore2015many,UnboundedGrowth,Colloquium,MBLinImpIsoQuaSys}.

One of the key points of the MBL state is the appearance of a large number of quasi-local integrals of motion (LIOMs), which limit the dynamics and prevent the cascade of energy across the system \cite{PhenomenologyFullMBL,Imbrie,LIOMs}. This means that highly excited many-body eigenstates have an area-law entanglement structure, as opposed to volume-law scaling of typical thermal states \cite{PhenomenologyFullMBL,UnboundedGrowth,Univslow,Colloquium}. These emergent conserved quantities give an insightful microscopic description of localization in interacting disordered systems and give rise to unconventional properties of the dynamics like no transport, long time memory retention and slow entanglement growth \cite{Univslow,PhenomenologyFullMBL,Imbrie,MBLandTherm}.

One of the defining dynamical features of MBL is the slow, logarithmic spread of entanglement. Weak interactions lead to exponentially small corrections to the energy of localized many-body eigenstates, which leads to slow dephasing between different many-body localized configurations. Although the system stays localized this interaction-induced dephasing leads to a universal time evolution of entanglement entropy with a logarithmic growth. This behavior is the result of the hierarchy of exponentially separated dephasing times that exists in interacting disordered systems, and thus the distant parts of the system do not become entangled until exponentially long times \cite{Univslow,UnboundedGrowth,Colloquium,PhenomenologyFullMBL,LIOMs}.

Even more complicated is the case of a periodically driven system described in terms of Floquet theory \cite{Bukov2015,FloquetMagnus,EquilibriumStatesofGeneric}. In this type of system, energy added to the system continuously from a periodic source; in these systems, generic interacting systems are expected to absorb energy indefinitely, and should eventually reach a kind of infinite-temperature equilibrium. In this regime, the initial conditions of the system were forgotten by the local observables, and the system goes to a featureless thermal state. This unlimited heating is a serious obstacle to the realization of stable non-equilibrium phases of driven quantum matter \cite{Effective,EnergyFlowinPeriodicTherm,Colloquium,FateofManyBody}.

Remarkably, it was shown that a strongly disordered system can inhibit this process of heating, thus the emergence of a Floquet MBL regime with non-ergodic behavior due to localization preventing the absorption of heat becomes feasible \cite{TMBlocalizationinperiodic,PerDriErandMBL}. The localization within periodically driven MBL systems happens in Floquet eigenstates; as a result, the system cannot reach the infinite temperature thermalization state. Moreover, at sufficiently high frequencies, the system's dynamics can be characterized by an effective quasi-Hamiltonian \cite{Effective,Expslow,Colloquium}.

Generic interacting periodically driven systems experience unlimited heating due to the continual input of energy from the periodic external drive and tend toward a high-temperature effective state. In this state, the local  forget about the initial conditions and behave similarly to a thermalized state \cite{Bukov2015,FateofManyBody,EquilibriumStatesofGeneric}. Periodically driven transverse-field Ising model analyzed in this paper and it is a good example of this heating phenomenon in nonintegrable spin chains. Here, the process of heating was described by analyzing the evolution of energy density and entanglement entropy upon each driving cycle \cite{Effective,EnergyFlowinPeriodicTherm,FloquetMagnus,Univslow}.

This sections outline the structure of the paper. In Section~\ref{M_TF}, we discuss about the effect of heating in periodically driven spin chain. We define the Hamiltonian as  clean and  with disorder. Then using Floquet formalism we find $H_{eff}$ and $K_{eff}$ and do the van Vleck inverse frequency expansion to get approximate values of $H_{eff}$ and $K_{eff}$. Using the ground state wavefunction of the approximate Hamiltonian and we calculate energy density and entanglement entropy density. Also we observe how it varies with driving frequency as the system is finite-sized. In Section~\ref{WoQD}, we analyze the stroboscopic dynamics of the system when its driving protocol is clean. We discuss the behavior of the energy density and entanglement entropy density curve when the system is exposed to intermediate/high, low and excessively high frequency regime. In Section~\ref{WQD}, we add quenched disorder to the parallel and transverse field component of the Hamiltonian. We analyze that how the system evolves and reaches Floquet-MBL regime with suitable driving frequency. Our observations are overall summarized as conclusion in Section~\ref{Conc}.

\section{Model and Theoretical Framework}\label{M_TF}

\subsection{Heating in disordered spin chains}
\label{2.1}

When an isolated spin chain is driven periodically, it absorbs energy from the external frequency driving which will lead to heating. We use the periodically-driven transverse-field Ising model with parallel field to study the energy density and entanglement entropy density \cite{PW}. As the Hamiltonian is time-dependent and non-integrable \cite{PW}. Therefore, it can be expressed as two-step protocol:
\begin{equation}
	\label{4eq}
	H(t) =
	\begin{cases} 
		\begin{aligned} 
			&J \sum_{j=0}^{L-1} \sigma_j^z \sigma_{j+1}^z  + h \sum_{j=0}^{L-1} \sigma_j^z, 
		\end{aligned} 
		& t \in \left[-\frac{T}{4}, \frac{T}{4}\right] \\[15pt] 
		
		g \sum_{j=0}^{L-1} \sigma_j^x, 
		& t \in \left[\frac{T}{4}, \frac{3T}{4}\right] 
	\end{cases} 
	\pmod{T},
\end{equation}
\begin{equation}
	\label{4}
	\begin{aligned}
		H(t) &= \frac{1}{2} \sum_{j=0}^{L-1} \left( J \sigma_j^z \sigma_{j+1}^z + h \sigma_j^z + g \sigma_j^x \right) \\
		&\quad + \frac{1}{2} \, \mathrm{sgn}[\cos(\Omega t)] \sum_{j=0}^{L-1} \left( J \sigma_j^z \sigma_{j+1}^z + h \sigma_j^z - g \sigma_j^x \right).
	\end{aligned}
\end{equation}

Where the $J$ represents the nearest-neighbor interaction strength, $g$ is the transverse field and $h$ is the parallel field with respect to the Ising axis.

Periodic driving in an isolated spin chain typically induces absorption of energy from the external driving which may ultimately lead to heating. When longitudinal quenched disorder ($h_j$) and transverse quenched disorder ($g_j$) are introduced in the clean Hamiltonian Eq. \ref{4eq}. The growth of energy and entanglement entropy can be strongly suppressed due to quenched disorders. These give rise to slow dynamics and localization like behavior \cite{PW}.

In order to observe these quantum effects we consider a periodically driven disordered spin-1/2 chain with periodic boundary conditions.  The system is now described by a time-dependent Hamiltonian in which nearest-neighbour interactions and on-site fields are subject to quenched disorder. The Hamiltonian takes the form as,
\begin{equation}
	\label{Actual}
	\begin{split}
		H(t) &= \frac{1}{2} \sum_{j=0}^{L-1} \left[ J \sigma_j^z \sigma_{j+1}^z + h_j \sigma_j^z + g_j \sigma_j^x \right] \\
		&\quad + \frac{1}{2} \mathrm{sgn}[\cos(\Omega t)] \sum_{j=0}^{L-1} \left[ J\sigma_j^z \sigma_{j+1}^z + h_j \sigma_j^z - g_j \sigma_j^x \right].
	\end{split}
\end{equation}

The on-site field $h_j$ is drawn from a uniform distribution $h_j \in [-W_h, W_h]$, where $W_h$ characterizes the disorder strength. The parameter $g$ denotes the transverse field and $g_j\in[-W_g,W_g]$ which denotes random transverse field disorder, while $\Omega$ is the driving frequency. The periodic driving enters through the signum function $\mathrm{sgn}[\cos(\Omega t)]$, which implements a two-step drive protocol. The presence of quenched disorder breaks translational and longitudinal invariance and plays a crucial role in suppressing energy absorption from the drive \cite{Effective,TMBlocalizationinperiodic}. As a result, the system exhibits slow heating dynamics and can enter a localization-dominated regime, which we analyze through the evolution of energy density and entanglement entropy.
\subsection{Floquet Formalism}\label{2.2}

The dynamics of a quantum system subject to a time-periodic Hamiltonian $H(t+T)=H(t)$ can be systematically analyzed using Floquet theory, which provides a natural framework to describe driven many-body systems \cite{FloquetMagnus,PW}. The time-evolution operator between two arbitrary times $t_1$ and $t_2$ can be expressed as,
\begin{equation}
	U(t_2, t_1) = \mathcal{T}_t \exp\left(-i \int_{t_1}^{t_2} H(t)\, dt \right)
\end{equation}
According to Floquet's theorem, the evolution operator can be factorized as,
\begin{equation}
	U(t_2, t_1) = e^{-iK(t_2)} e^{-iH_F(t_2 - t_1)} e^{iK(t_1)}
\end{equation}
where $H_F$ is the Floquet Hamiltonian and $K(t)=K(t+T)$ is a periodic kick operator encoding micromotion within a driving cycle. If we consider stroboscopic dynamics, then it can be given as,
\begin{equation}
	U(t_0 + nT, t_0) =  \exp\left(-i H_F[t_0]nT\right)
\end{equation}

In general, obtaining $H_F$ exactly is not feasible for interacting many-body systems. However, in the high-frequency regime  much larger than local energy scales, one can construct an effective Floquet Hamiltonian($H_{eff}$) using van Vleck inverse-frequency expansion \cite{Bukov2015,PW}. This expansion is used to calculate $H_{eff}$ in a different basis.
\begin{equation}
	\label{Heff}
	H_F=\exp[-iK_{eff}(0)]H_{eff}\exp[iK_{eff}(0)]
\end{equation}
The approximated terms can be written as, 
\begin{equation}
	\begin{split}
		H_F &= H_F^{(0)} + H_F^{(1)} + H_F^{(2)} + H_F^{(3)} + \mathcal{O}(\Omega^{-4}) \\
		&= H_F^{(0+1+2+3)} + \mathcal{O}(\Omega^{-4}).
	\end{split}
\end{equation}
\begin{equation}
	H_{\mathrm{eff}} = H_{\mathrm{eff}}^{(0)} + H_{\mathrm{eff}}^{(1)} + H_{\mathrm{eff}}^{(2)} + H_{\mathrm{eff}}^{(3)} + \mathcal{O}(\Omega^{-4})
\end{equation}
\begin{equation}
	K_{\mathrm{eff}} = K_{\mathrm{eff}}^{(0)} + K_{\mathrm{eff}}^{(1)} + K_{\mathrm{eff}}^{(2)} + K_{\mathrm{eff}}^{(3)} + \mathcal{O}(\Omega^{-4})
\end{equation}
We can write the effective Hamiltonian and the kick operator as,
\begin{equation}
	H_{\mathrm{eff}} = \sum_{n=0}^{\infty}{H_{\mathrm{eff}}^{(n)}}
	\hspace{0.5cm},
	\hspace{0.5cm}
	K_{\mathrm{eff}}(t) = \sum_{n=0}^{\infty}{K_{\mathrm{eff}}^{(n)}(t)}
\end{equation}
where $H_{\mathrm{eff}}^{(n)}$ and $K_{\mathrm{eff}}^{(n)}$ both are of the order of $\Omega^{-n}$. Now the Fourier decomposition of equation(\ref{Heff}) can be given like
\begin{equation}
	H_{\mathrm{eff}}^{(0)} = H_0= \frac{1}{T}\int_{0}^{T}H(t)dt
\end{equation}
\begin{equation}
	H_{\mathrm{eff}}^{(1)} = \frac{1}{\hbar\Omega}
	\sum_{l=1}^{\infty} \frac{1}{l} [H_{-l}, H_l]
\end{equation}
\begin{equation}
	\label{Heff2}
	\begin{split}
		H_{\mathrm{eff}}^{(2)} &= \frac{1}{\hbar^2\Omega^2} \sum_{l \neq 0} \Bigg( \frac{1}{2 l^2} [H_{-l}, [H_0, H_l]] \\
		&\quad + \sum_{\substack{l' \neq 0 \\ l' \neq l}} \frac{1}{3 l l'} [H_{-l'}, [H_{l'-l}, H_l]] \Bigg).
	\end{split}
\end{equation}
The expansion for the kick operators can be given as
\begin{equation}
	K_{\mathrm{eff}}^{(0)}(t)=0
\end{equation}
\begin{equation}
	K_{\mathrm{eff}}^{(1)}(t) = \frac{1}{i\hbar\Omega}\sum_{l \neq 0}\frac{\exp{(il\Omega t)}}{l} H_l
\end{equation}

\subsection{Observables}
\label{2.3}

To understand the dynamics of the periodically driven disordered spin chain we consider quantities that measure energy absorption, thermalization and quantum entanglement \cite{PW,nandkishore2015many,TwoBand}. Specifically we consider the energy density and the entanglement entropy that offer complementary perspectives on the dynamics of heating and localization. The energy density is a measure of the heating under periodic drive. It is the expectation value of effective (Floquet) Hamiltonian per site, 
\begin{equation}
	\epsilon(lT) = \frac{1}{L} \langle \psi_i | \exp{(ilTH_F)}H_{\mathrm{F}}^{(0+1+2)} \exp{(-ilTH_F)}| \psi_i \rangle
\end{equation}
Where L is the system size and $|\psi_i\rangle$ is the ground state of the approximate Hamiltonian $H_{F}^{(0+1+2+3)}$.
To probe the growth of quantum correlations, we compute entanglement entropy \cite{PW}. We consider a subsystem A and define it to contain L/2 consecutive chain sites:
\begin{equation}
	s_{ent}(lT)= -\frac{1}{L_A}tr_{A}[\rho_A(lT) log \rho_A(lT)],
\end{equation}
where, \hspace{0.5cm}
\begin{equation}
	\rho_A(lT)=tr_{A^c}[\exp{(-ilTH_F)}|\psi_i\rangle\langle\psi_i|\exp{(ilTH_F)}]
\end{equation}

$\rho_A$ is the reduced density matrix over the complement of A \cite{PW}. As we are considering many-body case, so we can split our system into two parts i.e. $\mathcal{L}$ and $\mathcal{R}$.The interaction energy leads to dephasing effect, and hence reduces off diagonal elements of density matrix, and thereby gives rise to entanglement. At a particular time t, only degrees of freedom up to a distance
\begin{equation}
	x(t) \sim \ln\left(\frac{t}{t_{\min}}\right)
\end{equation}
from the boundary between subsystems $\mathcal{L}$ and $\mathcal{R}$ are affected by dephasing \cite{Univslow}. The entanglement entropy can be expressed as diagonal entropy:
\begin{equation}
	S_{ent}(t)= CS_{diag} \hspace{0.5cm} S_{diag}=-\sum{P_i(x)lnP_i(x)}
\end{equation}
where $P_i(x)$ is the probability of the system being in a configuration $|\alpha\rangle$ of size $x$, derived from the wavefunction of the starting state. $S_{diag}$ is the diagonal entropy; this is essentially the highest possible entropy of a system in a particular starting state, assuming that the interactions do not change the eigenbasis.

For initial product states the diagonal entropy is proportioal to the subsystem size,
\begin{equation}
	S_{\mathrm{ent}}(t) \propto \xi \, \log\left(\frac{V t}{\hbar}\right).
\end{equation}
where $V$ is characteristic interaction strength causing dephasing between localized states \cite{Univslow,TwoBand}.
\subsection{Numerical Method}\label{2.4}
We study the dynamics of a disordered, periodically driven transverse field Ising model with parallel field using exact diagonalization \cite{PW,Bukov2015}. The computations are carried out in the entire Hilbert space of dimensions $2^L$ 
(without symmetry sectors) by utilizing the QuSpin package for building operators and their evolution. System sizes with $L\leq12-14$ are considered based on convergence and computational resources. The pure states evolve under the time-dependent Hamiltonian through exact time evolution which can be computed numerically by using the integrators implemented in QuSpin at discrete time intervals. The observables are measured stroboscopically at multiples of the driving period $T=2\pi/\Omega$ to minimize the influence of micromotion \cite{Bukov2015,FloquetMagnus}. 
 
\section{Stroboscopic dynamics for energy and entanglement entropy density without quenched disorder}\label{WoQD}
\begin{figure}[h]
	\centering
	\includegraphics[width=\linewidth]{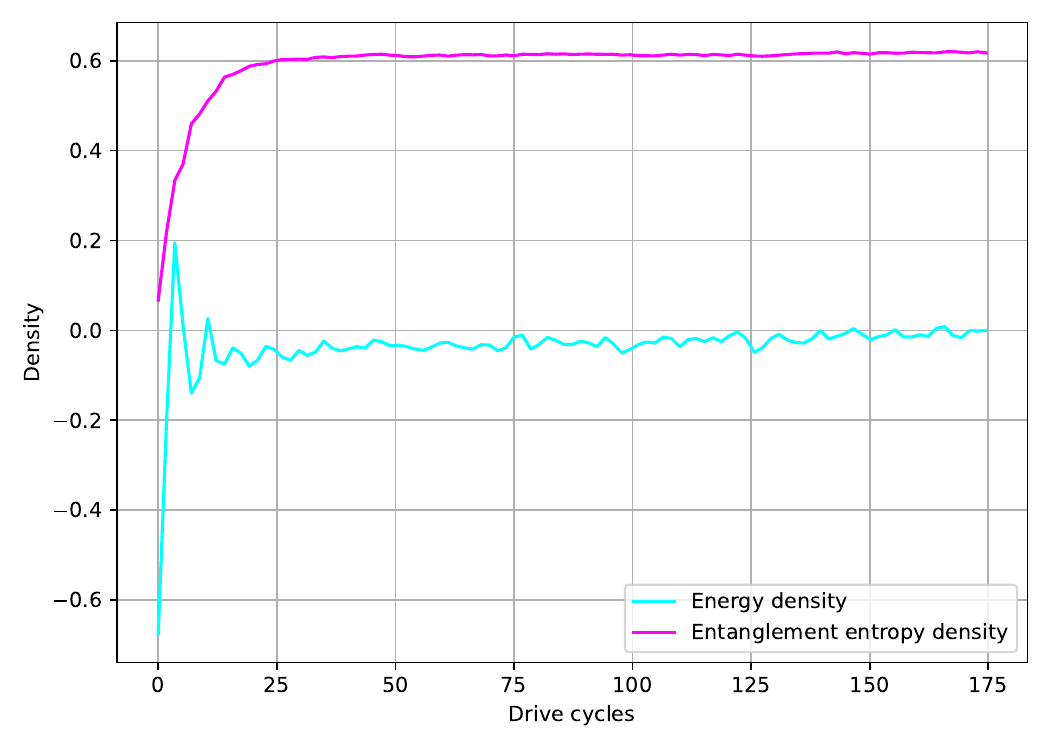}
	
	\caption{Stroboscopic dynamics of the energy density and entanglement entropy density together in a clean periodically driven TFIM in a parallel field. Parameters: $g/J$ = 0.915, $h/J$ = 1.236, $\Omega/J$ = 3.5998 and L = 14.}
	\label{3.1}
\end{figure}

The dynamics depicted in Fig. 1. represent the intermediate/high-frequency Floquet regime of the periodically driven transverse-field Ising chain with parallel field. We have used equation \eqref{4}. The parameters we have selected ($\Omega$ = 3.5998 $>$ $J, g, h$) mean that the driving frequency is greater than any of the intrinsic energy scales in the system, which leads to a significant suppression of rapid Floquet heating, allowing for the stabilization of an effective prethermal regime. The energy density shows bounded oscillations around an approximately constant average, without systematic drift toward infinite temperature, so energy absorption into the periodically-driven system occurs only weakly. This is typical of high-frequency Floquet dynamics, the system evolves nearly as if it were being driven by an effective static Floquet Hamiltonian that results from the Magnus or van Vleck expansion. The entanglement entropy density initially increases due to generation of quantum correlations through unitary evolution before eventually saturating at a finite value; the saturation is due to both finite size effects ($ L = 14$) and the limited Hilbert space available to the subsystem. Since the model is translationally invariant, strongly non-integrable and has no quenched disorder, the observed growth in entropy is not indicative of many-body localization, but rather characteristic of prethermalization in Floquet systems, with local observables relaxing to quasi-steady values, while heating is strongly suppressed over long time scales. The residual oscillations originate from finite-size effects and coherent interference between Floquet quasienergy modes, which persist because the finite system contains only a discrete quasienergy spectrum.

\begin{figure}[htbp]
	\centering
	\includegraphics[width=\linewidth]{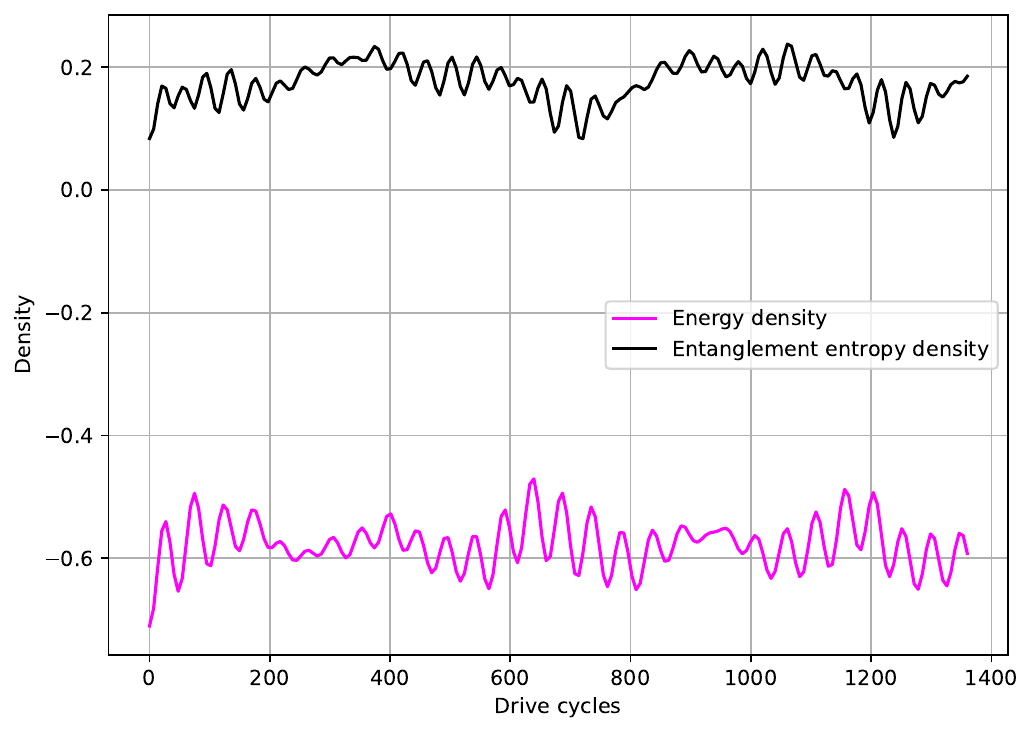}
	\caption{Stroboscopic dynamics of the energy density and entanglement entropy density together in a clean periodically driven TFIM in a parallel field. Parameters: $g/J$ = 1.2, $h/J$ = 0.02, $\Omega/J$ = 0.924 and L = 12.}
	\label{3.2}
\end{figure}

The fact that the entanglement entropy density approaches its maximal value also shows that the system is approaching an effective infinite-temperature Floquet state, in which local observables forget their initial conditions. While Floquet many-body localized systems are stabilized by strong disorder to prevent heating and remain non-ergodic, the present clean near-resonant system is efficiently thermalized by resonant Floquet heating.

In the Fig. 2. the dynamics is consistent with a clean interacting Floquet system driven in the low-frequency regime, which is nearly integrable in true sense but here the drive is sufficiently slow to remain near-resonant with the intrinsic many-body energy scales, such that it efficiently mixes a large number of many-body states in each cycle, but rather undergoes repeated non-perturbative transitions that lead to the observed large-amplitude oscillations in the energy density. The disorder-free nature of the system is key, as there is no mechanism for localization in energy space. 

\begin{figure}[htbp]
	\centering
	\includegraphics[width=\linewidth]{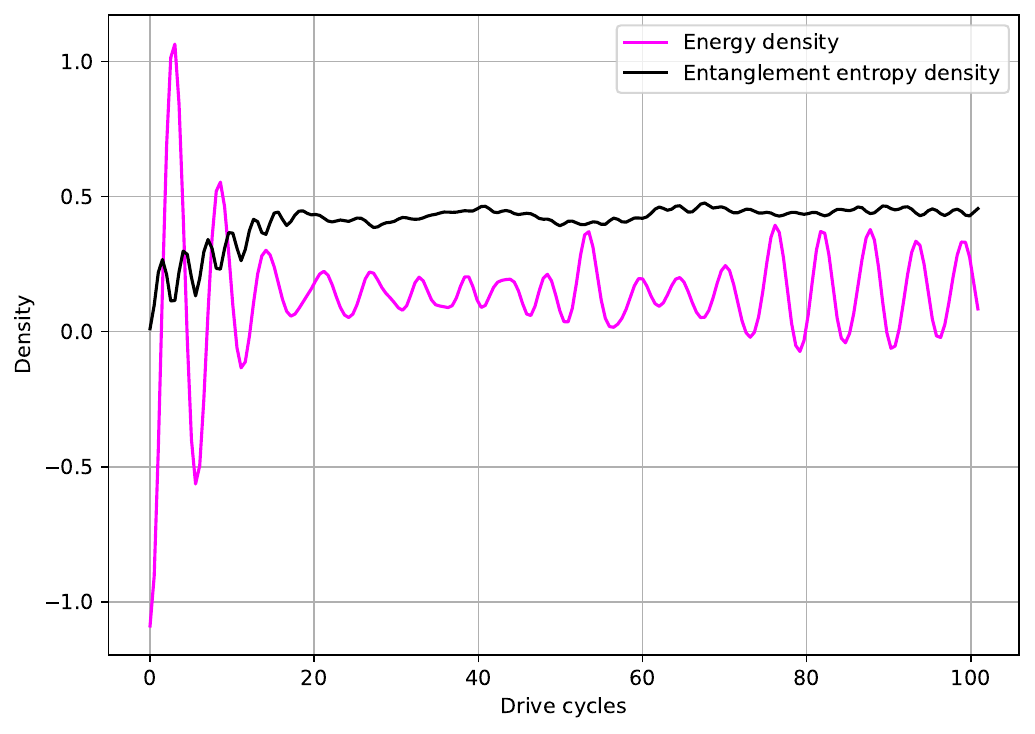}
	\caption{Stroboscopic dynamics of the energy density and entanglement entropy density together in a clean periodically driven TFIM in a parallel field. Parameters: $g/J$ = 1.2, $h/J$ = 2.536, $\Omega/J$ = 12.458 and L = 12.}
	\label{3.3}
\end{figure}

The entanglement entropy density is oscillating. This is because quantum correlations are being made and changed over the spin chain. The Floquet evolution is what makes the entanglement change as the wave function moves around a lot in the Hilbert space. We do not see the entanglement entropy getting stuck at a value. This means that the dynamics of the system is not constrained by many-body localization instead  the coherent Floquet mixing and finite size recurrences are governing. If we had a large or extended system and we kept driving it at a low frequency we would expect it to heat up a lot and eventually act like it is infinite temperature Floquet state. However for the time we can observe the system is still showing some behavior before it gets to that point and this behavior is oscillatory prethermal state. The entanglement entropy of the system is still changing, which is what we see when we look at the quantum correlations, in the chain. 

The dynamics shown in the Fig. 3. display the heating behavior of a periodically driven interacting spin system without quenched disorder, taking the driving frequency $\Omega$ much higher than the longitudinal field strength $h$. In this regime the external periodic drive($\Omega$) is much greater than the intrinsic energy scales($J$, $g$, $h$) of the system, leading to the suppression of energy absorption from the drive. This causes the energy density to be bounded and it exhibits only coherent oscillations rather than systematic drift toward the infinite temperature limit. Due to the bounded oscillations, the quantum phase information is preserved.

At the same time, initially the entanglement entropy density grows quickly with the number of drive cycles and saturates to a nearly steady-state value, suggesting the efficient spreading of quantum correlations across the system, a hallmark of finite size system and Floquet prethermalization dynamics. As there is no quenched disorder and the driving frequency is much larger, so  the dynamics is governed by effective static Hamiltonian which is obtained from high frequency Magnus or van Vleck inverse frequency expansion. Fig. 3. depicts that  the system shows coherent finite sized Floquet dynamics instead of many-body localization.
\section{Stroboscopic dynamics for energy and entanglement entropy density with quenched disorder}\label{WQD}
In the case of quenching, we add random field disorder to the parallel field component of the  Hamiltonian.   

Here we consider a periodically driven disordered spin chain Hamiltonian with quenched disorder in the parallel and transverse field. The Hamiltonian can be given as defined in \eqref{Actual}.

\begin{figure}[htbp]
	\centering
	\includegraphics[width=\linewidth]{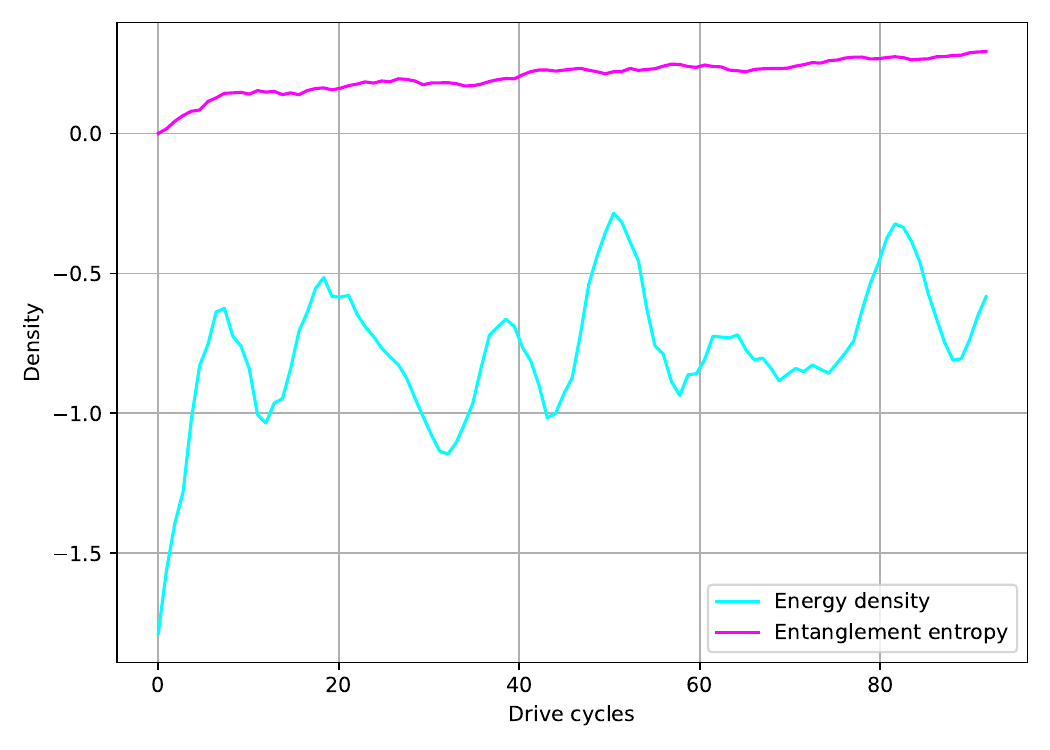}
	
	\caption{Stroboscopic dynamics of the energy density and entanglement entropy density together in the periodically driven TFIM in a parallel field with quenched disorder in parallel and transverse field components. Parameters: $g/J=0.915$, $h/J= 1.234$, $\Omega/J = 8.36$, $W_h= 5.85$, $W_g= 0.02 $ and L = 12.}
	\label{4.1.1}
\end{figure}

The result shown in the Fig.\ \ref{4.1.1}. shows the periodically driven disordered transverse-field Ising model with strong quenched disorder in the longitudinal field and are studied in the strong-disorder and high-frequency regime, where the disorder strength is much larger than the intrinsic interaction scales, and where signatures of Floquet many-body localization (Floquet-MBL) are clear.

The energy density is bounded and there is no systematic drift toward the infinite temperature limit. The graph shows that the energy density is not continuously growing upwards or downwards, rather it is oscillating around a nearly constant mean value. This indicates that the energy taken from the drive is strongly suppressed. 

Simultaneously the entanglement entropy density rises gradually and does not grow much which shows that quantum correlations stay in one place and do not spread easily across the chain. The large site-to-site energy mismatches associated with the strong random-field disorder($W_h/J =5.85$) hinder resonant spin rearrangements and the transport along the chain. The system therefore samples only a small subset of many-body Hilbert space, which results in the reduced entanglement production and in dynamics that are highly constrained. The small disorder due to bonds($W_J =0.02$) is negligible with respect to the dominant disorder due to the random field and does not significantly affect the localization properties.This kind of behavior is typical of many-body localization.In many-body localization disorder causes things to localize, stopping them from exploring all of Hilbert space in a way.

In the presence of strong quenched disorder and high frequency, the combined suppression of Floquet heating and the extremely slow growth of entanglement entropy provide strong evidence that the system remains in a stable Floquet-MBL phase. Consequently, the system retains memory of its initial state for long times and avoids thermalization despite the presence of periodic driving.


\section{Conclusion}\label{Conc} 
The non-equilibrium dynamics of periodically driven transverse field Ising chains have been explored in this work by following the dynamics of energy density and entanglement entropy density in the low frequency, high frequency and strongly disordered regimes. In the low frequency regime, the drive efficiently excites many-body excitations, which leads to higher resonant processes, more growth of entropy, and more tendency towards thermalization. In contrast, a high frequency driving gives rise to prethermal long-lived oscillating states of bounded energy density oscillations. With the addition of strong quenched disorder, the transport and spreading of quantum information are further suppressed, and a slow entanglement growth, Floquet-MBL like behavior is observed. In all parameter regimes, the residual oscillations are due to finite size effects and coherent interference of Floquet quasienergy modes. The results show that the driving frequency plays a key role in the interplay between interactions and disorder in controlling Floquet heating, prethermalization and localization phenomena and hence provide insight into the stabilization of long-lived non-equilibrium phases in periodically driven quantum many-body systems.
\begin{acknowledgments}
V.A. acknowledges support from the Science and Engineering Research Board (SERB), Anusandhan–National Research Foundation (ANRF), Government of India, through Core Research Grant No. CRG/2023/001573. Numerical calculations were performed using the PARAM Shavak (“Gryphon”) high-performance computing facility, whose support is gratefully acknowledged.
\end{acknowledgments}
\section*{data availability}

The data that support the findings of this article are available upon reasonable request.

\bibliography{ref}

\end{document}